# Re-calibrating methodologies in social media research:
*Challenge the visual, work with Speech*

*H. Kevin Jin*
kevin.jin@student.uva.nl

*Abstract*

This article methodologically reflects on how social media scholars can effectively engage with speech-based data in their analyses. While contemporary media studies have embraced textual, visual, and relational data, the aural dimension remained comparatively under-explored. Building on the notion of secondary orality and rejection towards purely visual culture, the paper argues that considering voice and speech at scale enriches our understanding of multimodal digital content. The paper presents the *TikTok Subtitles Toolkit* that offers accessible speech processing readily compatible with existing workflows. In doing so, it opens new avenues for large-scale inquiries that blend quantitative insights with qualitative precision. Two illustrative cases highlight both opportunities and limitations of speech research: while genres like *#storytime* on TikTok benefit from the exploration of spoken narratives, nonverbal or music-driven content may not yield significant insights using speech data. The article encourages researchers to integrate aural exploration thoughtfully to *complement e*xisting methods, rather than replacing them. I conclude that the expansion of our methodological repertoire enables richer interpretations of platformised content, and our capacity to unpack digital cultures as they become increasingly multimodal.

Table of Contents

# INTRODUCTION

Contemporary media studies has been driven by a central goal: to understand how practices, discourses, and cultural forms emerge, diffuse, and intersect in our reality through evolving and increasingly complex media technologies (Valdivia 2003, 3). Today's social media platforms offers a wealth of cultural activity and *locus* for exploration, yet this abundance also derives methodological challenges. The very notion of *content* has expanded: we came to know content as singular media objects: text, image, or sound; each carries symbols and meanings of their own. With emerging digital technologies and their embodiments (such as social media platforms), content has encapsulated more modalities and combined previously individual media objects into one. An X/Twitter thread may contain both text and images; a TikTok clip assembles text, moving images, audio into one. As the notion expands, knowing *what content really means* becomes increasingly demanding. By complicating our understanding in its composition, we arrive at an epistemological challenge in unpacking the *assemblage* of meaning in new media content (Parry 2023): *how precisely* do we study content? What can we infer from them?

I use the term "study" to broadly allude to an arsenal of ever-expanding methodologies used by new media scholars, as few are exempt from interpreting content *at some scale*. Methods that originated in traditional media studies (close reading, discourse analysis, and content analysis) have been incorporated with computational approaches that embrace "natively digital" media (Özkula, Omena, and Gajjala 2024; Rogers 2024). These methods enable scholars to visit media content at scale, through patterns in hashtags, metrics (like, subscribe, view counts), and relations (reply and follow networks) in order to trace topic dissemination, community mobilisation, and vernacular patterns. Yet, aforementioned methods resolve little *the analytical impasse* put forth on social media scholarship, of accurate, comprehensive approaches suitable for smaller samples against the sheer *volume* and *velocity* of digital media, inherent to platformised activities that demands computational and automated research methods (Quan-Haase and Sloan 2016). Many scholars have favoured large-scale quantitative or mixed-methods analysis, sourcing data primarily from existing collections or APIs that are accessible and ready-to-use (cf., Hagen and Venturini 2024; Rogers 2021). Yet such approaches are not perfect: some data simply cannot be collected because they are restricted or unavailable; some study re-lies on human experiences and interactions which could not be performed at scale. As a result, much of the existing computational methodologies can be accused of lacking analytical precision and nuance. Focusing too closely on *certain* data risks overlooking others; and as I will illustrate, meaningful receptive dimensions of multimodal content, such as *the audible,* often fall victim in current research paradigms. The lack of discussion on speech and auditory experiences despite its prominence in extant social media literature can be partly attributed to the difficulty of collecting, processing and analysing them in large quantities, which then push researchers towards a path of less resistance using accessible data. As a result, audio and speech analysis has been primarily reserved for small-sample quantitive analysis rather than other stages of the research. Nevertheless, this paper aims to demonstrate that optimising our methodological inventory for aural media allows voice and speech to be studied *at scale* and *up close* to inform scholars *throughout their research* beyond interpretative analysis, while providing some much-needed *quality* in handling quantitative datasets.

## ON SPEECH, AND ITS ABSENCE IN SOCIAL MEDIA METHODOLOGIES

Research into speech and its mediation long predate social media. In light of emerging sound production and mediation technologies, scholars spoke of *acousmatic sound*[1], *acoustic space*[2], *anthrophony*[3] and *secondary orality* to conceptualise the various ways sound is employed and experienced across mediums and environments. Amongst them, *secondary orality* emerges as a particularly relevant concept to bridge the role of speech in social media content from its historical iterations. Building on McLuhan's "global village" concept that emphasised the spatial compression enabled by electric media, Ong delved deeper into the temporal dynamics instated by the digital, arguing that the emergence of telephone, radio, and television has created a sense of immediacy and ephemerality in media content similar to primary orality (Venturini 2022, 63). Unlike its *primary* form pre-dating literacy, *secondary orality* is a "more deliberate and self-conscious" form of oral-like quality inherently grounded upon textual (ibid., 68), lasting and replicable medium that could manifest in media that are not exclusively aural. Scholars have employed secondary orality to analyse various media forms, including radio programs (Larson 1995; Angel 2014), television (Sen 1994), as well as audiovisual podcasts (Johansson 2021), in how they foster participation, with a remote audience of unknown identities and quantities (Angel 2014, 58). It has also been contested in non-aural online environments to theorise culture formations and community practices and where e*phemerality is intrinsic and attention is scarce*, such as 4chan (Hagen and Venturini 2024). Yet, despite extant research on secondary orality and various media that have established the significance of spoken language in literate and technologically saturated societies, in its "generating of a strong group sense, the concentration on the present moment, and its *formulaicness"* (Holly 2011, 353, highlight added), studies on the aural aspect of social media content have remained surprisingly scarce.

---

[1] coined by French composer Pierre Schaeffer; adopted in film studies by Michel Chion (2001). For more, see Kane (2014); Noudelmann (2018).

[2] this notion is perhaps most renowned in works of McLuhan and E. Carpenter, in distinction from the visual space. For an introduction, see Schafer (2007).

[3] a terminology for human-generated sounds in acoustic ecology; see Pijanowski et al. (2011) for an introduction specific to environmental humanities.

As introduced earlier, scholars have indeed curated methods and conventions for social media: Richard Rogers' "digital methods" considers web platforms as both a data source and a site of study, contextualising content analysis with networked and relational data (hashtags, user networks, links) (Rogers 2018; 2024); Lev Manovich's pioneering cultural analytics used large-scale image analysis to reveal aesthetic trends, cultural differences, and chronological progression on sites like Instagram and the TIME magazine (Manovich 2017; 2021). More recent studies have nuanced these approaches to accommodate evolving platforms with ethnographic work, computational processing and contextualising networks of hashtags, sounds and co-creation options (cf., Geboers and Van De Wiele 2020; Krutrok 2021; Schellewald 2021; 2023). While these methods have surely advanced the field, the auditory dimension of audiovisual content equally prevailing in this social media epoch has been less attended to in comparison.

Analysing speech can be indeed challenging, for speech has to be collected, transcribed, parsed and interpreted; all of which demands additional effort and expertise. Much of the extant social media "field work" has been criticised to follow a path of least resistance, focusing on what scholars have critiqued as "easy" data (Burgess and Bruns 2015; Özkula, Reilly, and Hayes 2023) from platforms that are more accessible in their acquisition and usage, sourcing quantifiable metrics, classifiable metadata and images using research-friendly tools and APIs. However, easy data is not all that there is: Manovich prompted scholars to consider "dimensions and aspects of culture that existing measurements do not capture" (2020, 10); Özkula and colleagues (2024) further warned that "easy" data research risks insufficient insights. As hashtags and replies represent communities and extracted texts and visuals stand isolated from their coalescing modalities[4], hard-to-capture cultural manifests and activities remain out of the peripheral. The inadequacy of easy data does not cease here, as a narrow scope of modality omits the fact that media is simultaneously experienced otherwise. On a philosophical note, W.J.T. Mitchell rejects the notion of purely visual media or culture, as "there is no such thing as pure visual perception in the first place" (2005, 264); even those appearing so, such as silent films, inherently engage with multiple senses, such as touch and hearing. As such, sensory interplay is fundamen-tal to the experience and understanding of any media, and conditions how media "enters into the region of emotion, affect and intersubjective encounters" (ibid., 264). Moving on to social media, scholars should note that no amount of hashtags and visual cues represent the content itself; nor will they fully illuminate narrative strategies, emotional registers, or cultural subtleties encoded in the spoken word. A TikTok video might picture a mundane activity, such as dressing up, but meaningfully shape the content with voice-overs that resists conventional visual- and metadata-based methods. *It is precisely here* that speech offers methodological and analytical merits that may further our understanding on content and its planetary practices. Its analytical lineage should remind us that *the voice* warrants analysis as both a meaningful instrument *producers* wield and a provocative factor in *consumption*.

---

[4] For example, the omission of speech when one extracts only visuals from video content; or the neglect of interactive affordances (such as Facebook reactions) that could alter how one reads content. For the latter, see Geboers et al. (2020).

In practice, studying speech in multimodal content, such as TikTok videos, can be approached quantitatively *and* qualitatively. While some *qualitative* analysis may be possible without data processing, auxiliary procedures such as transcription render speech *legible* for textual analysis and accessible for scaled computational methods such as sentiment analysis, topic modelling, and other natural language processing (NLP) techniques. Such macro-level analyses can reveal emergent themes, narrative patterns, and affective dynamics *uttered* across large corpus of content. Qualitatively, discourse analysis of specific scripts may locate identity performances, engagement strategies and participatory culture in action. Speech *in itself and through transcription* could then synergise with existing methodologies media researchers deployed for visual, textual and relational data in how spoken discourse reinforces, challenges, or complements other dimensions of content.

In the following sections, I will empirically outline some benefits and limitations for studying speech in social media content through my explorations on TikTok. Alongside my exploration, a novel toolkit for accessible spoken-text mining on TikTok is presented[5], as well as exemplar protocols that familiarise researchers with the tool.

---

[5] The Toolkit is openly available at github.com/j-nivekk/miscdataworks/tree/2d10c8d93224b8630583c6578f35f0fad46d8f00/TikTok

**INCORPORATING SPEECH IN TIKTOK CONTENT ANALYSIS**

THE TIKTOK SUBTITLES TOOLKIT

One of the first obstacles researchers face in studying speech and audio content emerge from its feasibility. Unlike "easy" data, speech and transcribed text are usually not available through platform APIs and existing datasets. Despite developments in digital research tools such as *4CAT*, *Zeeschuimer* and *Voyant Tools*, collecting and analysing speech in multimodal social media content at scale typically requires speech-to-text processing, a resource-intensive and technically demanding task that oftentimes discourages humanities scholars. Building on the canonical social media scraping tool *Zeeschuimer* (Peeters 2024), the *TikTok subtitles toolkit* offers an alternative route for speech text mining that significantly reduces local speech-to-text processing. It achieves so by locating and scraping platform-side transcription stored as *subtitles* in datasets captured via *Zeeschuimer*. This approach substantially lowers the barrier to incorporating speech into data-driven analyses, allowing scholars to consider *orality* at scale. Additionally, the Toolkit is built with compatibility in mind to streamline with 4CAT both in its input and output. Its output can be customised to embed in source datasets for further processing. Scholars can thereby correlate speech data with information about creators, hashtags, engagement metrics, or the use of effects/stickers to *contextualise* and *interpret* content, whilst addressing prior analytical inadequacies related to visual and network driven methodologies.

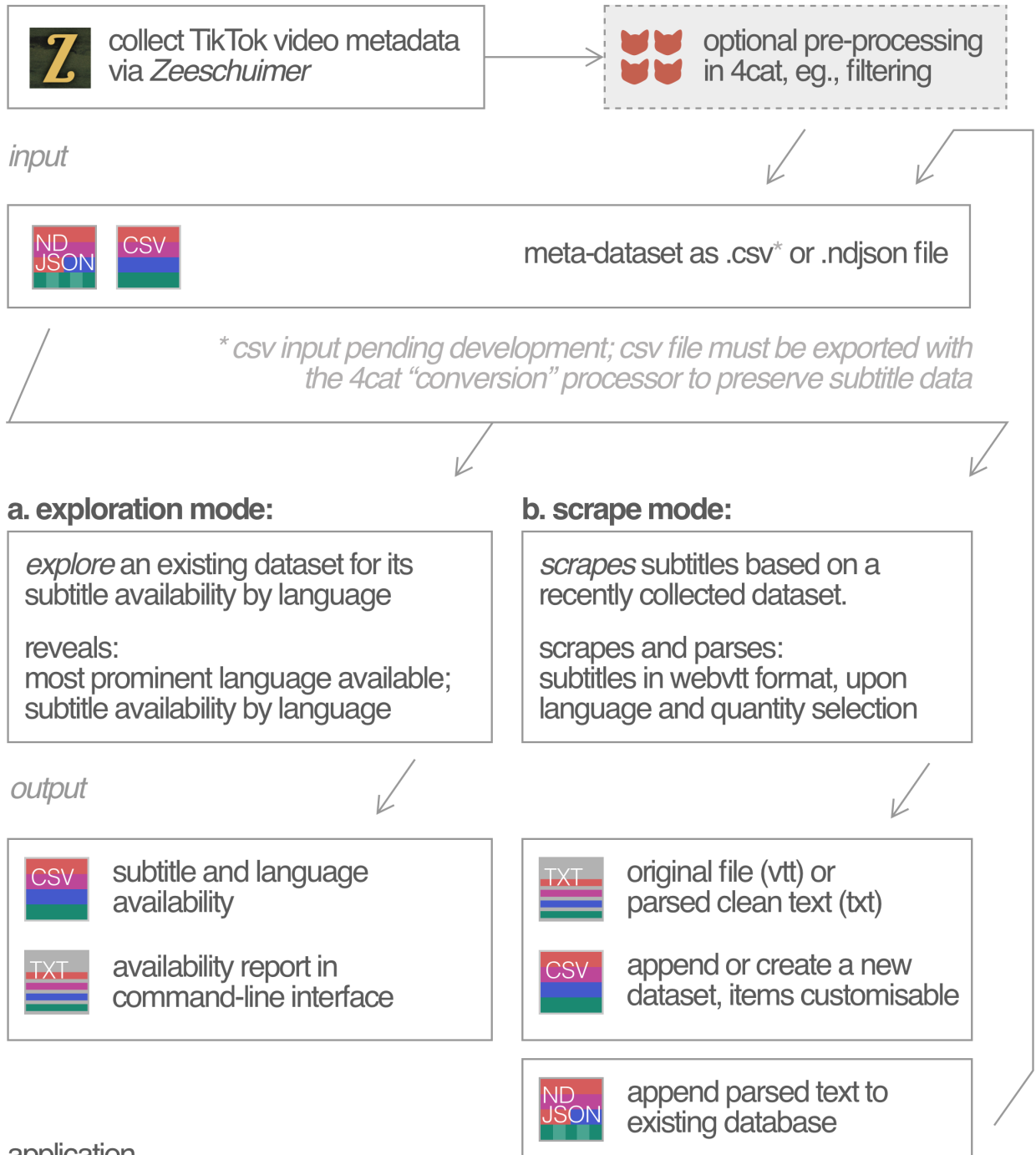

## TOOLKIT IN ACTION: OPPORTUNITIES AND LIMITS

As with any methodological advancement, studying speech in applying the *Toolkit* requires careful consideration of the research question, the nature of the dataset, and the communities under study. Here, I report on two exploratory cases: one in which speech analysis proves useful, and another that highlights the limitations of approaching orality.

Our first case samples content under the *#storytime* hashtag, canonically used by creators to share stories. On TikTok, one soon notes that this genre consistently uses speech and narration in its delivery. Researchers, then, may explore the language or themes employed by individual creators or communities. At a larger scale, questions may arise in whether specific topics receive more engagement than others, or whether storytelling as a *practice* exhibits certain characteristics specific to TikTok. Metadata and visual cues are insufficient in answering such questions for they omit the *auditory* (**Figure 1**), whereas traditional speech-to-text approaches would *at least* involve sampling, scraping and audio processing that require extensive time and computing resources. Meanwhile, the Toolkit is able to retrieve hundreds of transcripts within seconds. While not all speech data may be collected through our Tool, it significantly reduces the processing workload required to quantitatively examine a genre that is *vocal.* The lightweight processing of the Toolkit thus incentivises speech research as it forgives dispersive explorations that would otherwise be tedious to proceed: yet when speech *corpora* can be generated at negligible costs, researchers may rapidly develop research questions by identifying outliers and trends in speech length, vocabulary and theme through text analysis tools such as *Voyant Tools*.

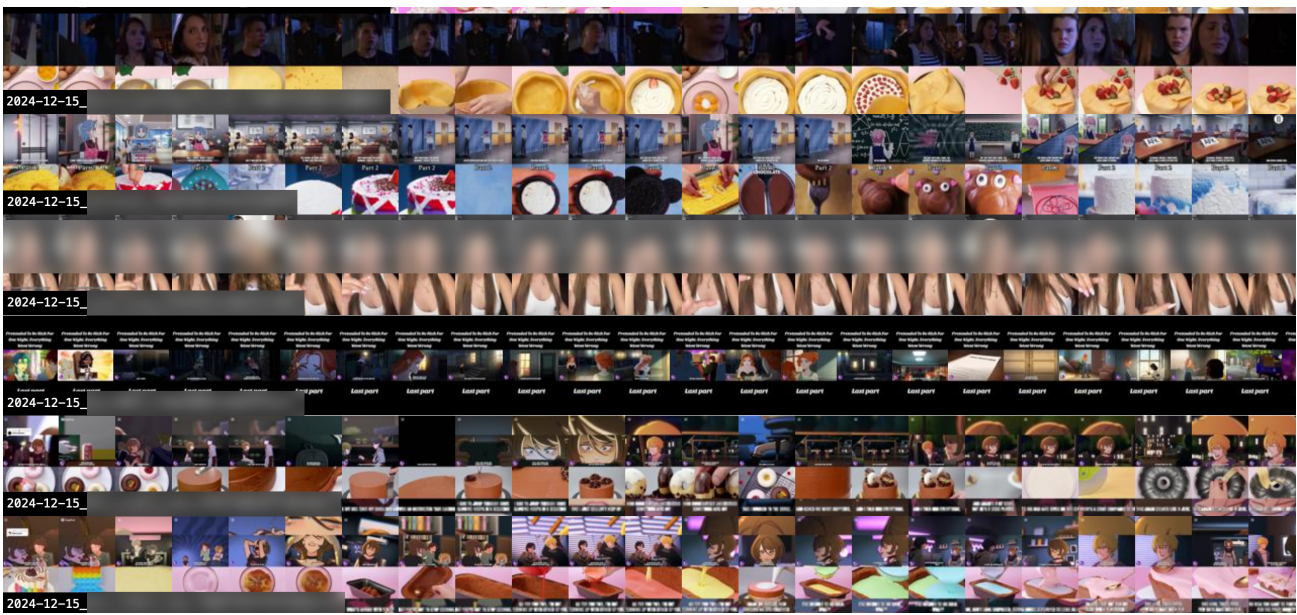

**Figure 1.** Scene-by-scene timelines of *#storytelling* content: aural storytelling cannot be captured through visual cues. Generated with 4CAT.

Our next case builds on the work by Geboers and colleagues (Forthcoming) on TikTok and war that queries content by the sound "*Доброго вечора Where Are You From*". The authors noted that the nonverbal sound can meaningfully contribute to the content while serving as an nexus for networked multimodal engagement beyond the auditory, such as visual presentation and editing styles. This is confirmed in my scraping exploration using the sound, where speech data is evidently scarce (**Figure 2**). In this example, visual approaches in conjunction with hashtag networks would yield more meaningful results than speech analysis. Additionally, the presence of deaf or signing communities, and genres that eschew spoken word such as dance or music, should remind scholars that speech analysis is not a panacea to substantiate all social media research. Instead, the absence of speech can become a subject of inquiry: the use of nonverbal sounds or alternative modalities, such as captions are deliberate aesthetic and communicative choices that may reflect particular subcultures, accessibility conditions, or platform affordances. An overemphasis on speech in such cases would then misconstrue the nature of the community's engagement and creativity; as such, it is essential to observe the specificities of an issue space and its vernaculars before dedicating towards speech analysis at scale.

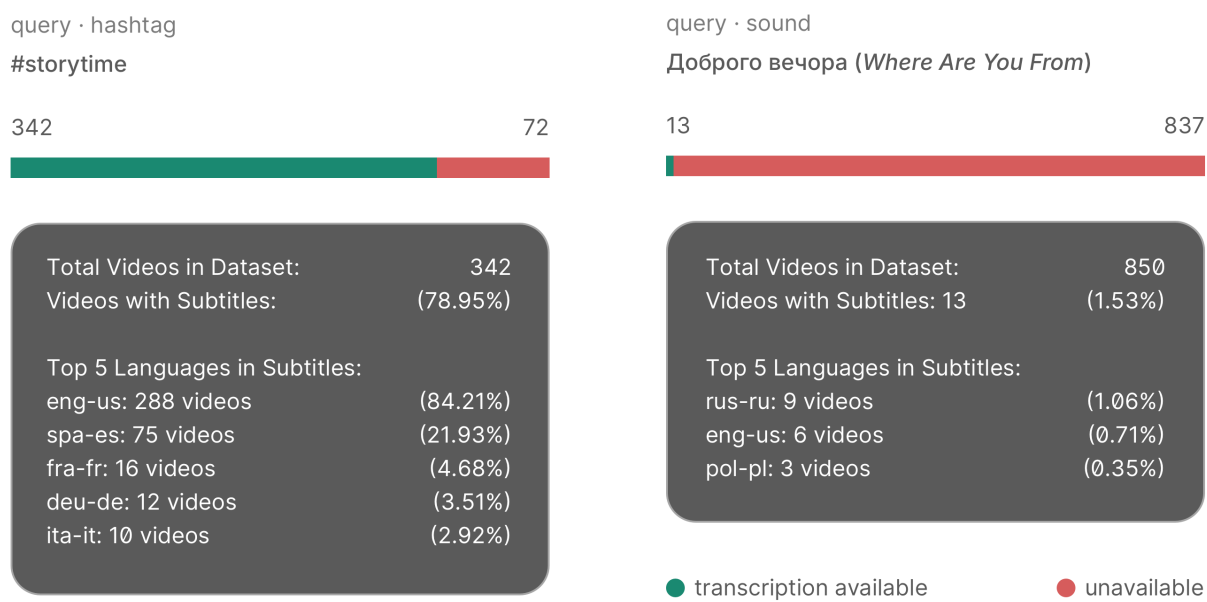

**Figure 2.** Occurrence of speech data by query; the effectiveness of speech analysis depends on the theme and community in question. Data generated by *Subtitles Toolkit.*

## Conclusion

Historical scholarship on radio, television and podcasts have aptly demonstrated the vital role of voice and speech across mediums, whereas Ong's *secondary orality* further clarified why sound and voice matters in technically saturated societies. As we acknowledge the manifold sensory modalities of digital content, studying speech is a conceptual turn to the fundamental understanding of the *body* as an apparatus of communication, "a zone of resonance" and an interface of perception, which has been somewhat sidelined in the digital era due to practical difficulty and conceptual omission (Stahl 2013, 89; Burgess and Bruns 2015). From here, the presented *Subtitles Toolkit* facilitates the integration of speech capture into existing computational workflows and provides an accessible means to operationalise this insight.

At the same time, the researcher must remain attentive to the contexts of their study, recognising *when speech analysis is relevant* and when *other modalities or methods* better serve their research question. With speech data also comes new ethical considerations, given the personal and often vulnerable nature of spoken content on social media platforms. Researchers must respect individual privacy and diligently interpret users' utterances to not exploit or distort the voices they study. More data does not translate to better findings, yet refining our methodological inventory propels new curiosities and insights. In the epoch of platformised social media, methodologies embracing the growing complexity of content have indeed been long due.

13## NOTES

Datasets discussed in this paper may contain sensitive information. As such, they are only available upon contact with the author for inquiries.

Source code and documentation of the *Subtitles Toolkit* presented in this paper are available at GitHub: https://github.com/j-nivekk.

15Ong, Walter J. 2009. *Orality and Literacy: The Technologizing of the Word*. Reprinted. New Accents. London: Routledge.

Özkula, Suay M., Paul J. Reilly, and Jenny Hayes. 2023. 'Easy Data, Same Old Platforms? A Systematic Review of Digital Activism Methodologies'. *Information, Communication & Society* 26 (7): 1470–89. https://doi.org/10.1080/1369118X.2021.2013918.

Özkula, Suay Melisa, Janna Joceli Omena, and Radhika Gajjala. 2024. 'Researching Visual Protest and Politics with "Extra-Hard" Data'. *Journal of Digital Social Research* 6 (2): 46–65. https://doi.org/10.33621/jdsr.v6i2.214.

Parry, Kyle. 2023. *A Theory of Assembly: From Museums to Memes*. Minneapolis: University of Minnesota Press.

Peeters, Stijn. 2024. 'Zeeschuimer'. Zenodo. https://doi.org/10.5281/ZENODO.14418239.

Peeters, Stijn, and Sal Hagen. 2022. 'The 4CAT Capture and Analysis Toolkit: A Modular Tool for Transparent and Traceable Social Media Research'. *Computational Communication Research* 4 (2): 571–89. https://doi.org/10.5117/CCR2022.2.007.HAGE.

Quan-Haase, Anabel, and Luke Sloan. 2016. 'Introduction to the Handbook of Social Media Research Methods: Goals, Challenges and Innovations'. In *The SAGE Handbook of Social Media Research Methods*, by Luke Sloan and Anabel Quan-Haase, 1–9. 1 Oliver's Yard, 55 City Road London EC1Y 1SP: SAGE Publications Ltd. https://doi.org/10.4135/9781473983847.n1.

Rogers, Richard. 2021. 'Visual Media Analysis for Instagram and Other Online Platforms'. *Big Data & Society* 8 (1): 20539517211022370. https://doi.org/10.1177/20539517211022370.

———. 2024. *Doing Digital Methods*. 2nd ed. SAGE Publications Ltd. http://gen.lib.rus.ec/book/index.php?md5=1E7D8ED03A1E4D9985B12B1950D8EA3F.

Schellewald, Andreas. 2021. 'Communicative Forms on TikTok: Perspectives From Digital Ethnography'. *International Journal of Communication* 15 (0): 21.

———. 2023. 'Understanding the Popularity and Affordances of TikTok through User Experiences'. *Media, Culture & Society* 45 (8): 1568–82. https://doi.org/10.1177/01634437221144562.

Sen, Nabaneeta Deb. 1994. 'Secondary Orality: How Words Speak Through Television'. *Indian Literature* 37 (5 (163)): 125–41.

Stahl, Heiner. 2013. 'Sound Studies. An Emerging Perspective in Media and Communication Studies'. In *Past, Future and Change. Contemporary Analysis of Evolving Media Scapes.*, edited by Ilija Tomanic Trivundza, N. Belakova, Iris Jennes, Şafak Dikmen, Joanna Kedra, Sander De Ridder, and Giulia Airaghi. https://www.academia.edu/6635713/Past_future_and_change_Contemporary_analysis_of_evolving_media_scapes.

Valdivia, Angharad N., ed. 2003. *A Companion to Media Studies*. Blackwell Companions in Cultural Studies 6. Oxford: Blackwell. https://doi.org/10.1002/9780470999066.

Venturini, Tommaso. 2022. 'Online Conspiracy Theories, Digital Platforms and Secondary Orality: Toward a Sociology of Online Monsters'. *Theory, Culture & Society* 39 (5): 61–80. https://doi.org/10.1177/02632764211070962.